\documentclass{article}
\usepackage[utf8]{inputenc}
\usepackage{amsmath,amsfonts}
\usepackage{graphicx}
\usepackage{lineno}
\usepackage{xcolor}

\usepackage{xifthen}
\usepackage{lineno}
\newboolean{shortarticle}
\newboolean{displaycopyright}
\setboolean{shortarticle}{true}
\setboolean{displaycopyright}{true}

\def\curl{\ensuremath{\mathbf{curl}}\,}

\def\bE{\ensuremath{\mathbf{E}}}

\def\bJ{\ensuremath{\mathbf{J}}}
\def\be{\ensuremath{\mathbf{e}}}
\def\bh{\ensuremath{\mathbf{h}}}

\def\bx{\ensuremath{\mathbf{r}}}
\def\bx{\ensuremath{\mathbf{x}}}

\def\tensmurd{\ensuremath\boldsymbol{\mu}_{r,1D}}
\def\tensmurdd{\ensuremath\boldsymbol{\mu}_{r,2D}}

\def\tensepsrd{\ensuremath\boldsymbol{\varepsilon}_{r,1D}}
\def\tensepsrdd{\ensuremath\boldsymbol{\varepsilon}_{r,2D}}

\def\om{\ensuremath{\omega}}

\newcommand{\gui}[1]{\textcolor{black}{#1}}

\def\bE{\ensuremath{\mathbf{E}}}
\def\bH{\ensuremath{\mathbf{H}}}

\def\bEinc{\mathrm{\mathbf{E}}^{\mathrm{inc}}}

\def\bEd{\mathrm{\mathbf{E}}^{\mathrm{d}}}

\def\bEtot{\mathrm{\mathbf{E}}^{\mathrm{tot}}}

\title{Complete design  of a fully integrated graphene-based compact plasmon coupler for the infrared}
\newboolean{oldstyle}
\setboolean{oldstyle}{false} 
\ifthenelse{\boolean{oldstyle}}{%
\author[1]{Aswani Natarajan} 
\author[1]{Guillaume Demésy}
\author[1,*]{Gilles Renversez}
\affil[1]{Aix Marseille Univ, CNRS, Centrale Marseille, Institut Fresnel, 13013 Marseille, France}
\affil[*]{Corresponding author: gilles.renversez@univ-amu.fr}
\begin{abstract}
 A fully integrated waveguide-based, efficient surface plasmon coupler composed of a realistic non-tapered dielectric waveguide with graphene patches and sheet is designed and optimized  for the infrared. The coupling efficiency can  reach nearly 80\% for a coupler as short as 700 nm for an operating wavelength of 12 $\mu$m. This \gui{work} is carried out \gui{using rigorous numerical models} based on the finite element method \gui{taking into account 2D-materials as surface conductivities}. \gui{The} key numerical results are \gui{supported by} physical arguments based on modal approach or resonance condition.
\end{abstract}
}%
{\author{Aswani Natarajan, Guillaume Demésy,  and Gilles Renversez\\
Aix Marseille Univ, CNRS, Centrale Marseille, Institut Fresnel,\\Marseille, France\\
gilles.renversez@univ-amu.fr}
}
\begin{document}
\graphicspath{{./}}
\maketitle
\begin{abstract}
  A fully integrated waveguide-based, efficient surface plasmon coupler composed of a realistic non-tapered dielectric waveguide with graphene patches and sheet is designed and optimized  for the infrared. The coupling efficiency can  reach nearly 80\% for a coupler as short as 700 nm for an operating wavelength of 12 $\mu$m. This \gui{work} is carried out \gui{using rigorous numerical models} based on the finite element method \gui{taking into account 2D-materials as surface conductivities}. \gui{The} key numerical results are \gui{supported by} phdisplaycopyrightysical arguments based on modal approach or resonance condition.
\end{abstract}
Since the renewal of plasmonics in the last two decades, the generation and launching of surface plasmon polaritons (SPPs) has been a crucial problem both theoretically and experimentally~\cite{Ditlbacher:08,PARK20094513,Tetienne:11,doi:10.1021/nl302040e}. 
Potential applications of SPPs include sensing and integrated photonics. 
In fully integrated configurations based on optical waveguides, useful to reach compactness and robustness for future photonic devices, one of the most efficient solutions that has been proposed is to use a metal grating to ensure the generation of SPPs from the input beam. The metal grating allows to compensate the mistmatch between the  propagation constant   of the mode of the input fully dielectric waveguide and the  propagation constant  of the mode that propagates in the output waveguide that is  of plasmonic type due to the presence of an usually thin metal layer.
This concept of coupler to generate  SPPs has already been proven experimentally in the mid-infrared near 8 $\mu$m~\cite{PhysRevLett.104.226806}.
\gui{Our goal} is to design such optical device but for the longer wavelengths of the mid-infrared \gui{and} typically we choose 12 $\mu$m. But at this wavelength, the ratio of the real and imaginary parts of the metal permittivity is less favourable for SPP generation than the one obtained at shorter wavelengths. One solution to overcome this limitation is to consider graphene~\cite{PhysRevB.80.245435,2010NaPhovol4p611B,doi:10.1021/nn406627u} due to its peculiar conductivity. Another key feature of graphene is its tunalibity~\cite{2011NatNanotechvol6p630J}. For example, its complex conductivity can be largely modified using a voltage bias across the 2D-material ensuring a direct, rapid, adjustable control of the graphene properties and consequently of the coupler. It must be pointed out that \gui{even} if in the present work only graphene is considered, the \gui{present} method can be used to tackle other 2D-materials like  hexagonal boron nitride sheet or silicen~\cite{doi:10.1021/cr300263a}\gui{: the results obtained} can be generalized for these materials when their electromagnetic properties can be modeled by 2D conductivities.    
The \gui{issues} linked to the use of graphene arise from the fact that the graphene plasmons (GPs) have much higher propagation constants than \gui{those} of SPPs and consequently that the \gui{required} phase matching condition \gui{implies a} smaller grating period.

In order to model the full device accurately and rigorously, \gui{we adapted our recent method~\cite{Demesy20}} to study discontinuities in waveguides within a full vector description given by Maxwell’s equations. It doesn’t rely on any hypothesis regarding \gui{sizes, shapes or permittivities} of the discontinuities, and it doesn’t require \gui{any approximation} as long \gui{as linear} materials are considered. In the present study, the graphene \gui{inclusions form} the discontinuities. \gui{The incident} mode of the input \gui{waveguide} is computed within the framework of the finite element method (FEM) in a modal approach. It is then injected as an incident field in the device containing the grating in a scattered field approach, again within the FEM framework. The grating is followed by a continuous graphene sheet where the GP can propagate. All the physical quantities, either local or global, like the Poynting vector or the coupling efficiency can be computed accuretaly. Graphene is modelled as a genuine 2D sheet and described by a scalar complex conductivity $\sigma_{gr}$ given by the Kubo’s model~\cite{Li-Tao-Chen-Huang:16}. This accurate \gui{2D} way to describe the graphene has at least two advantages compared to the more \gui{conventional} way where a \gui{layer with finite thickness} is artificially introduced to describe \gui{graphene}. First, it avoids the study of the dependency of the results as a function of the artificial layer thickness (See Supplement 1). Second, in 3D problems, it avoids to model a finite thickness layer with 3D elements since only 2D elements are needed to mesh the graphene sheet, which substantially reduces the computational resources. In 2D models like the one investigated here, it avoids to model a thick layer with 2D elements. The FEM formalism we used to solve the 2D problem and its implementation using the open source softwares gmsh and getdp~\cite{gmsh,getdp} \gui{are} also valid to tackle \gui{3D problems}.

In order to design a realistic coupler that can work at 12~$\mu m$, the guiding layer is made of the $\mathrm{Te}_{20}\mathrm{As}_{30}\mathrm{Se}_{50}$ (TAS) chalcogenide glass due to its high transparency at this wavelength~\cite{desevedavy09tas-mof}. The \gui{grating coupler} (see Fig.~1) is made of $N$ identical graphene patches with a period $\Lambda$ and a duty cycle $l_g/\Lambda$. It is longitudinally separated from the graphene sheet by a space $d$.

In this work, it is shown that a graphene-based coupler region as short as 700 nm, made \gui{of a few identical patches of graphene}, can provide a high conversion efficiency towards a highly localized GP around the final graphene sheet. The organization of the \gui{letter} is the following.  First, the theoretical framework to evaluate the electromagnetic fields is provided focusing on the differences with the works that are already available. Second, a general definition of the coupling efficiency \gui{is given} to evaluate quantitatively the different \gui{configurations.} Third, two examples of the field profiles along the device are shown in order to illustrate the method and to clarify the link between the fields and the coupling efficiency. Fourth, the general results for the coupler are given  as a function of the main parameters including the waveguide core thickness, the duty cycle of the diffraction grating, and the number of grating periods. 
These results are analyzed in terms of mode coupling in the different section of the structure, and also in terms of resonance condition. 
\begin{figure}[t] 
  \centering
  \includegraphics[width=0.9\linewidth]{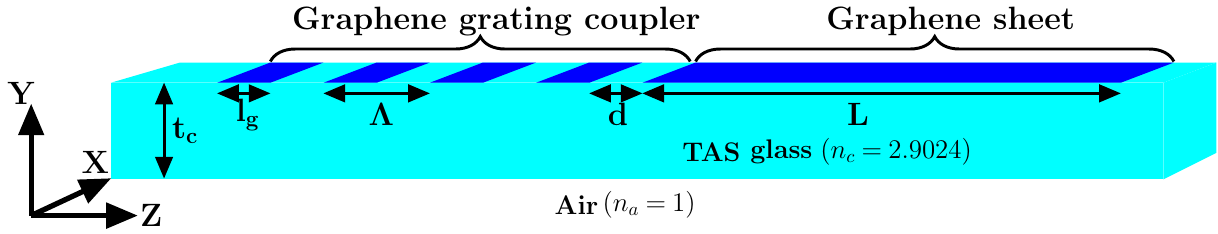} 
  \caption{Scheme of the graphene-based grating coupler around the coupling region. The diffraction grating made of identical patches of graphene (dark blue) is located on the top of the core waveguide  (cyan) before the continuous graphene sheet (dark blue)  located on the right part.}
  \label{fig:scheme-grating-coupler}
\end{figure}
In the investigated device (see Fig.~\ref{fig:scheme-grating-coupler}), it is assumed that its initial part (leftmost one) is invariant  toward the negative $z$ in order to make meaningful the use of a modal approach to determine the incident modes that can propagate along this axis in this fully dielectric region of the structure also assumed to be invariant along the $x$ axis. This classical 1D problem is solved introducing the ansatz $\bE =\be^{\mathrm{D}}_m(y)e^{-i(\om_0 t-\beta_{\mathrm{D},m} z)}$ where $\om_0$ is a given real angular frequency, $\beta_{\mathrm{D},m}$ the complex propagation constant   and $m$ is the index of the considered dielectric mode. The equation to solve in 1D is:
\begin{equation}\label{eq:Helmholtz}
  \curl\left(\tensmurd^{-1}\,\curl\bE\right) = \tensepsrd\left(\frac{\om_0}{c}\right)^2 \bE .
\end{equation}
It is worth mentioning that the  permeabilities and permittivities given in \eqref{eq:Helmholtz} are tensors due to the use of perfect matched layers in the FEM implementation used to solve the problem, and that the subscript ${1D}$ is a reminder of the assumed invariance of the corresponding quantities~\cite{livre12-FPCF}. \gui{Throughout} the study, all the physical materials are assumed to be nonmagnetic and the wavelength is set to 12 $\mu$m where the TAS refractive index of the waveguide core is set to 2.9024 as measured in Ref.~\cite{Carcreff21}. The explicit form of the corresponding quadratic non-Hermitian eigenvalue problem and more details on the FEM implementation are given in Ref.~\cite{Demesy20} while its weak form formulation can be found in Ref.~\cite{livre12-FPCF}. 

After extrusion of the 1D structure and \gui{breaking the $z$-invariance with the graphene patches}, one obtains $\tensepsrdd$ and  $\tensmurdd$. The previously computed eigenmodes can now be used as incident \gui{fields} $\bEinc$ on the coupler part of the device that will determine  $\bEtot$, the total field solution of the scattering structure made of \gui{both the waveguide and the graphene inclusions}. In order to treat the graphene elements as conductivity surfaces~\cite{PhysRevB.84.161407}, one needs to introduce the \gui{surface} current, non-null only on graphene, $\bJ_{gr}(\bx) =\sigma_{gr} \bEtot(\bx)$~\cite{Li-Tao-Chen-Huang:16} in the electromagnetic problem:  
\begin{equation}\label{eq:Helmholtz-general}
   \curl\left( \curl\bEtot\right) = \tensepsrdd\left(\frac{\om_0}{c}\right)^2 \bEtot   + i \om_0 \mu_0 \bJ_{gr}
\end{equation}
 For all the given results,  the following graphene parameters of the Kubo's model are used to evaluate $\sigma_{gr}$: Fermi energy $E_{f}$ = 0.6 eV, relaxation time $\tau$ = 0.5 ps, and temperature $ T$ = 300 K.

The outgoing scattered field  defined by $\bEd \equiv \bEtot-\bEinc$ can now be introduced and is the solution of the following equation obtained from the linearity of Eqs.~(\ref{eq:Helmholtz}-\ref{eq:Helmholtz-general}):
\begin{equation}\label{eq:Hdddrad}
  -\curl\left[\tensmurdd^{-1}\,\curl \bEd \right] + k_0^2\,\tensepsrdd\,\bEd = - i  \om_0 \mu_0 \sigma_{gr} \bEinc.
\end{equation}
This new 2D problem is no longer an eigenvalue problem and the surface localized source term \gui{proportional to} $ \bEinc$ in this equation can be evaluated straightforwardly from the chosen incident fields derived from the initial modal 1D problem. \gui{Note that in presence of dielectric inclusions as scatterers ($\tensepsrdd\neq\tensepsrd$), a bulk source term would also appear~\cite{Demesy20}.} The next step is to define the coupling efficiency $\eta$  toward the GP mode knowing the total field induced by the \gui{incident} one. 
 First, let us sum up the different modes involved in the full structure  that will be used in this definition and in the result section.
 The first kind of modes is the GP one of the structure covered by the graphene sheet, assumed to be invariant along the $z$-axis, and described by the electromagnetic fields 
 $\bE^{\mathrm{GP}}_{\mathrm{q}}= \be^{\mathrm{GP}}_{\mathrm{q}}(x,y)e^{-i(\om_0 t-\beta_{\mathrm{GP}} z)}$
 and  $\bH^{\mathrm{GP}}_{\mathrm{q}}= \bh^{\mathrm{GP}}_{\mathrm{q}}(x,y)e^{-i(\om_0 t- \beta_{\mathrm{GP}} z)}$ where  $ \beta_{\mathrm{GP}}$ is the GP propagation constant and $\mathrm{q}$ is the index of the considered GP mode. The second kind of modes is the hybrid one for the same structure that are typically waveguide core localized modes with a small part on the graphene sheet. They are described by the electromagnetic fields $\bE^{\mathrm{H}}_{\mathrm{p}}= \be^{\mathrm{H}}_{\mathrm{p}}(x,y)e^{-i(\om_0 t-\beta_{\mathrm{H}} z)}$ and $ \bH^{\mathrm{H}}_{\mathrm{p}}= \bh^{\mathrm{H}}_{\mathrm{p}}(x,y)e^{-i(\om_0 t - \beta_{\mathrm{H}} z)}$  where   $ \beta_{\mathrm{H}}$ is the hybrid mode propagation constant and  $\mathrm{p}$ is the index of the considered hybrid mode. 
 The definition of $\eta$ can be obtained in the following way:
 \setlength{\arraycolsep}{0pt}
\begin{multline}
  \label{eq:definition-eta-efficiency}
\eta =|C| \, \mbox{ with }  C=\int_{\substack{\;\\\Gamma_S}} [\bEtot\times{\bh^{\mathrm{GP}}_{\mathrm{q}}}    ].\mathbf{n} dS / \\ \Big( \sqrt{ \int_{\substack{\;\\\Gamma_S}} [\be^{\mathrm{D}}_{\mathrm{m}}  \times {\bh^{\mathrm{D}}_{\mathrm{m}}} ].\mathbf{n} dS } \sqrt{  \int_{\substack{\;\\\Gamma_S}}  [\be^{\mathrm{GP}}_{\mathrm{q}}  \times {\bh^{\mathrm{GP}}_{\mathrm{q}}} ].\mathbf{n} dS } \Big) %
\end{multline}
In the following, only the main GP modes and the main hybrid mode are considered (See Supplement 1).
 The integrals are computed on  $\Gamma_S$  which is a full cross-section of the studied structure, perpendicular to its main axis. These  $\Gamma_S$  cuts can be \gui{performed} at different $z$-location to follow the evolution of the efficiency along \gui{the device.} The denominators are used to normalize the two field terms of the numerator as it is explained in Ref.~\cite{sammut1976leaky,snyder1983optical} when leaky modes and  bi-orthogonality are involved. 
\begin{figure}[htbp!]
  \begin{center}
    \includegraphics[width=0.9\linewidth]{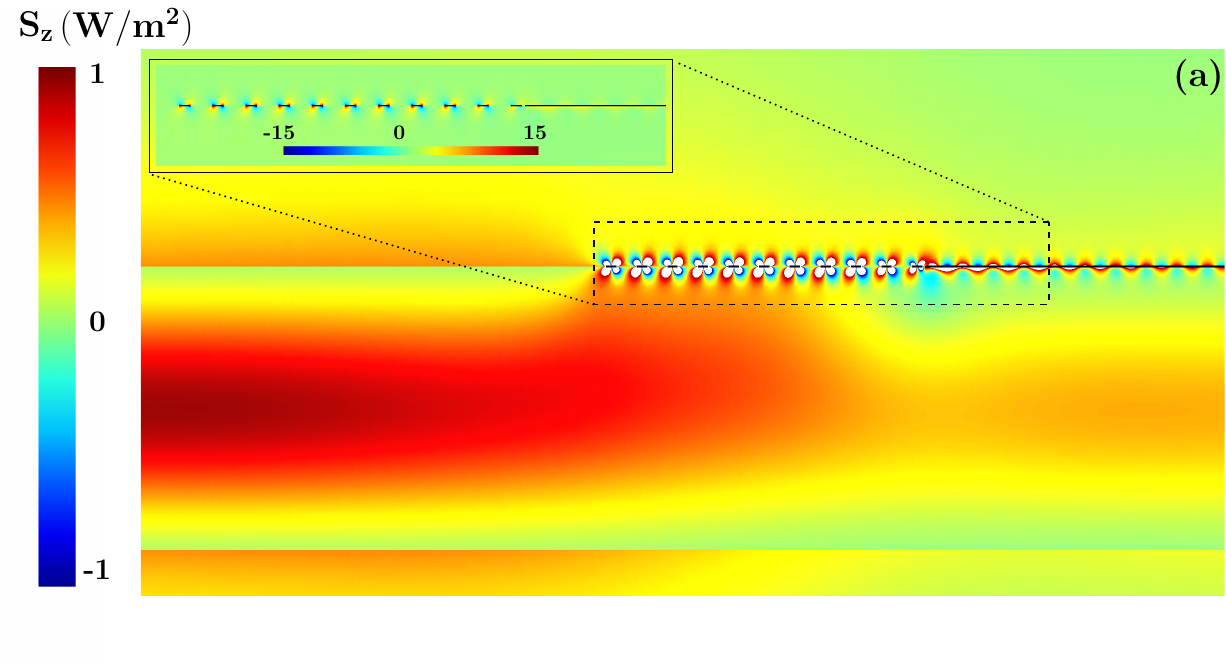}
    \includegraphics[height=0.45\linewidth,width=0.9\linewidth,clip=true,trim= 0 0 0 0]{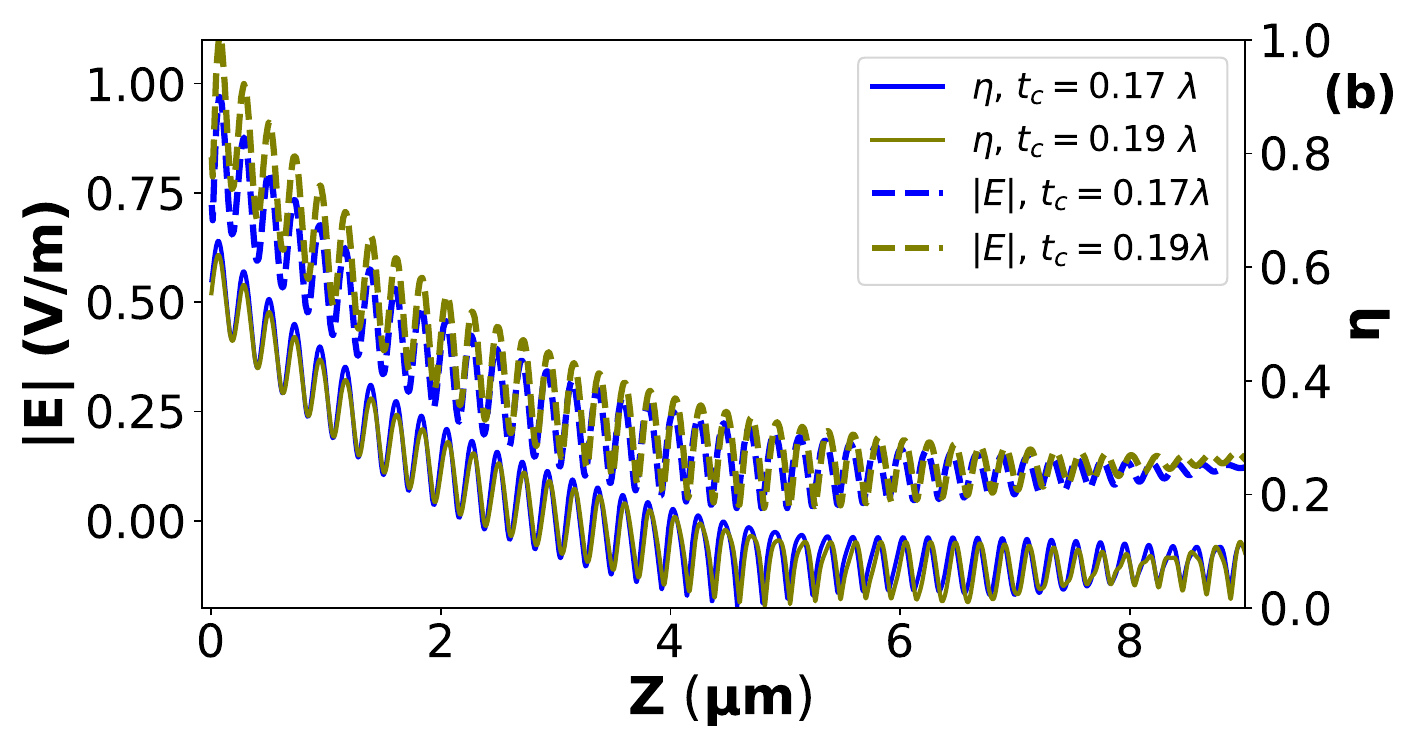}
    \caption{(a): $z$-component of the Poynting vector for $t_{c}$ = $0.17 \lambda$,  $\Lambda=220 $ nm, $d$ = 20 nm, $l_g/\Lambda= 0.3402$, $N$= 11. The top inset is a zoom of the coupler region where one can clearly see the individual graphene patches. (b): $|E_{tot}|$ and $\eta$ from $z$ = 0 (beginning of graphene sheet) to $z$ = 9 $\mu$m for $t_c =0.17\lambda$ and $t_c =  0.19 \lambda $ for $N=9$, $l_g/\Lambda=0.33575$, and $d=20$ nm.}\label{fig:field-plot-short-regime}\label{fig:link-field-Etot-eta} 
  \end{center}
\vspace{-0.5cm}
\end{figure}

The general phase matching condition between the incident dielectric D mode  and the GP mode within the diffraction grating of period $\Lambda$ is $\beta_{\mathrm{GP}} = \beta_{\mathrm{D}} + k \, 2 \pi/\Lambda$ with  $k \in \mathbb{Z}$~\cite{Raether88}. A typical value of $\Lambda$ is approximatively 220 nm for the studied configurations depending on the exact waveguide core thickness that has an impact on $ \beta_{\mathrm{D}}$. 
When \gui{analysing the} diffracted and total fields along the waweguide part covered with the graphene \gui{sheet,} one can typically distinguish two kinds \gui{of coupling phenomena}: \gui{a short range} one where the coupled field on the graphene sheet from the grating decays rapidly \gui{and a long range one} where the field on the graphene sheet comes directly from the  mode propagating in the core because the field from the grating has already faded. Our study is mainly dedicated to the first regime since it allows the design of compact and efficient coupler as shown in Fig.~\ref{fig:field-plot-short-regime}. The conversion from a guided core-localized mode on the left side of the structure to a configuration with both an hybrid mode and a GP one on the right side \gui{(where the graphene sheet is located)} is clearly illustrated in the top map of Fig.~\ref{fig:field-plot-short-regime}. The bottom graph of this figure illustrates the $z$-dependency of  $\bEtot$  along the graphene sheet after the coupler, and \gui{its link with} $\eta$. \gui{The} total electric field  and $\eta$ \gui{both oscillate} along the $z$-axis of the device.  The initial oscillations (typically before 10 $\mu$m) are generated by the beating between the  GP mode and the hybrid mode. These results can be quantified using Fourier transform analysis of the field cut along the $z$-axis where the main peaks are linked to the beating wavelength $\lambda_{beat} = 2 \pi / \Re e(\beta_1-\beta_2)$ between mode 1 and mode 2 (See Supplement 1). The observed  exponential decay of the field \gui{directly corresponds to the imaginary part} of the GP mode propagation constant that is much larger than the one of the hybrid mode.

\begin{figure}[htbp!]
  \begin{center}
    \includegraphics[height=0.3\linewidth,width=0.49\linewidth]{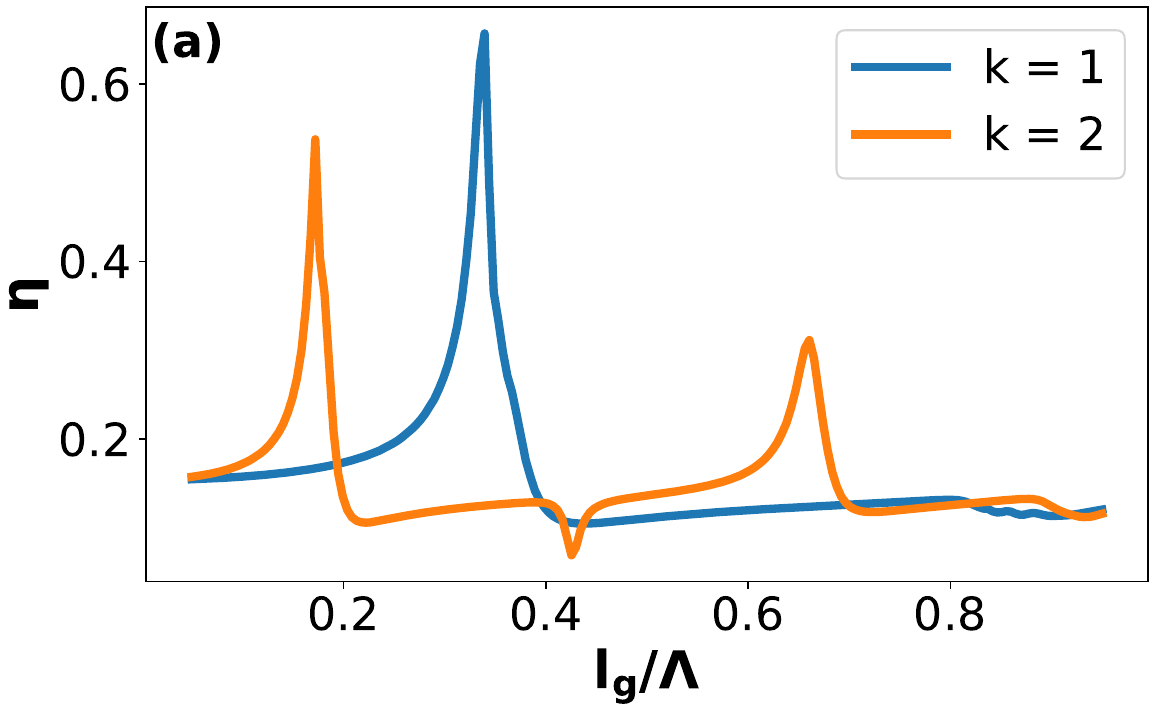}
    \includegraphics[height=0.3\linewidth,width=0.49\linewidth]{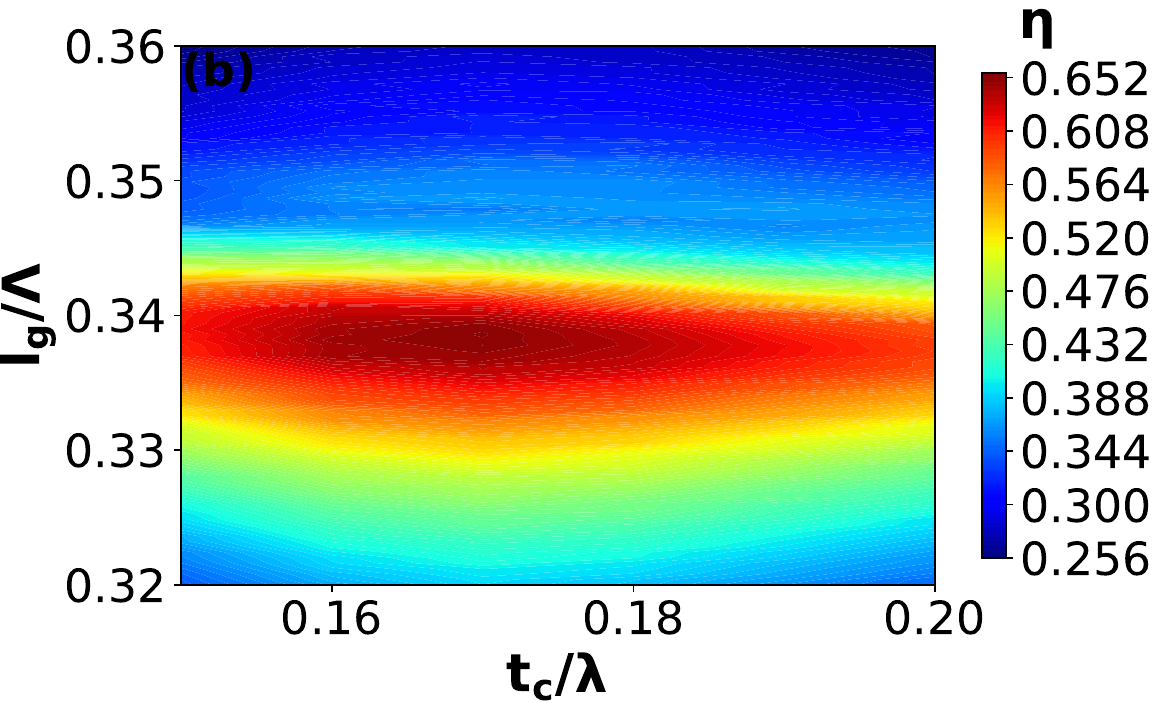}
    \caption{(a):  Coupling coefficient $\eta$ as a function of the duty cycle  $l_g/\Lambda$ for the first two diffraction orders for a core thickness $t_c/\lambda=0.17$ and  $N=11$. (b): Coupling coefficient $\eta$ as a function of the duty cycle  $l_g/\Lambda$ and of the core thickness $t_c$ for $N=12$ and $d=20$ nm.}\label{fig:duty-cycle-short-regime} 
  \end{center}
\vspace{-0.5cm}
\end{figure}

The duty cycle of the diffraction gratings is another key parameter~\cite{livre-brun-petit80}. In our case, it is defined by the ratio  $l_g/\Lambda$ and its impact is studied in Fig.~\ref{fig:duty-cycle-short-regime}. In the left part of this figure, the coupling coefficient as a function of the duty cycle is given for the first and the second orders $k$ of the phase matching condition. For $k=1$, a single peak centered around 1/3 is observed while for $k=2$ two peaks are present. Since the $k=1$ peak is higher and broader, this configuration is selected for the study. The theoretical optimal value of 1/3 for the duty cycle  can be explained from a multiple resonance \gui{condition,} simplified using the fact that for the studied structures $\Re e(\beta_D) \ll \Re e(\beta_{\mathrm{GP}})$, the position of two peaks obtained for $k=2$ can also be evaluated in the same framework (see Supplement 1). One can notice that the single peak duty cycle value is different from the one obtained for metal-based plasmonic coupler where values in the range $[0.45,\,0.6]$ are usually used~\cite{Tetienne:11}, again the middle of this interval can be approximated using a resonance condition depending on the metal permittivities (see Supplement 1).
The next key parameter that needs to be studied is the waveguide core thickness $t_c$. Its impact is monitored using the coupling coefficient $\eta$,  in the right part of Fig.~\ref{fig:duty-cycle-short-regime},  as a function of $t_c$ and of  $l_g/\Lambda$ for a value of N set to 12. It can be seen that $\eta$ is maximal for $t_c/\lambda$ in the interval [0.16,0.18] and that the best value for the duty cycle is 0.34 at least for $N=12$. In this range, a peak value of 65\% is reached for $\eta$ at the beginning of the graphene sheet.
The $N$ value has been chosen in order to ensure a sufficient number of periods for the grating but keeping the coupler length below 3 $\mu$m. As it will be shown later, even larger coupling coefficients can be reached in shorter couplers when the optimization process is pushed forward. Actually, the optimal \gui{$[0.16\lambda,0.18\lambda]$ range for the waveguide thickness} can be obtained using  the following argument within a simple modal approach. The coupling coefficient is enhanced when the electric field along the graphene is strong, and more precisely when its longitudinal components take high values since they are only the components that interact with the graphene. As it can be seen in Fig.~S1 in Supplement 1, the electric field single longitudinal component $E_z$ of the fundamental mode of the planar waveguide is maximal for a core thickness of 0.17 $\lambda$. This \gui{behavior} can be used in order to define \textit{a priori} a good approximation of the optimal waveguide core thickness of graphene couplers made of other materials than the ones used here even if, as shown later in this work, \gui{pure numerical studies} can improve further the coupler design.

\begin{figure}[htpb!]
  \begin{center}
    \includegraphics[width=0.8\linewidth]{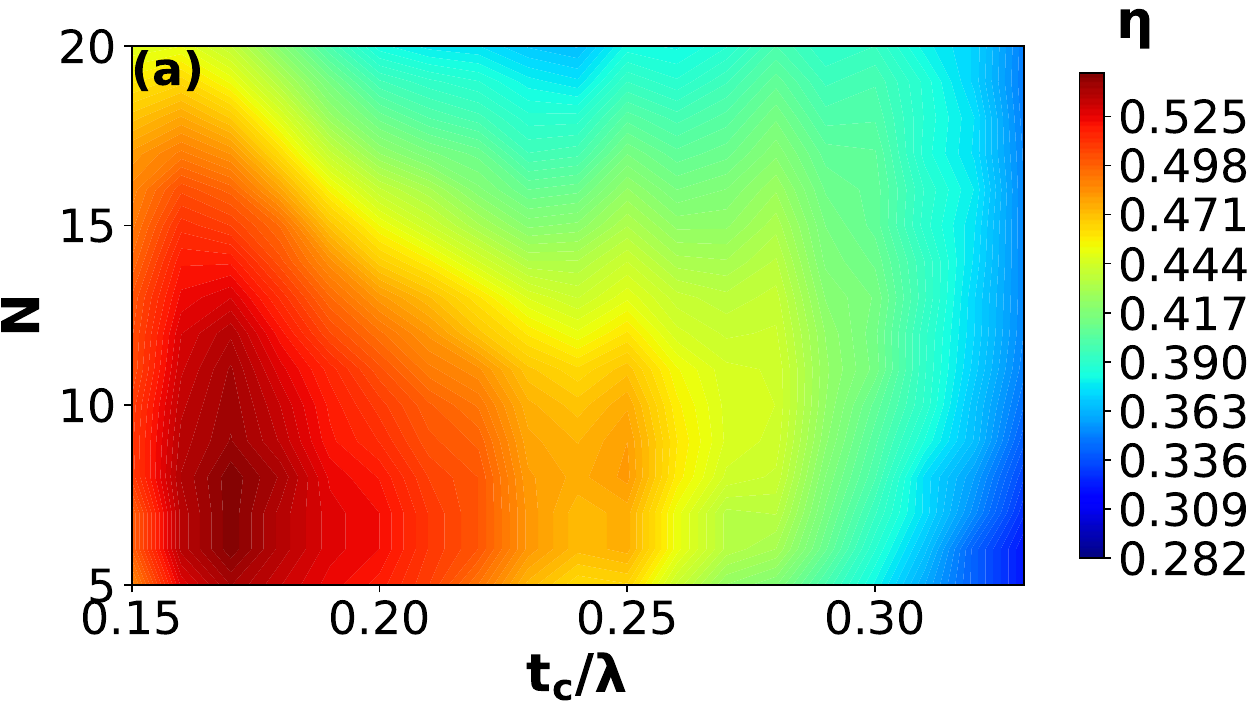}
    \includegraphics[width=0.8\linewidth]{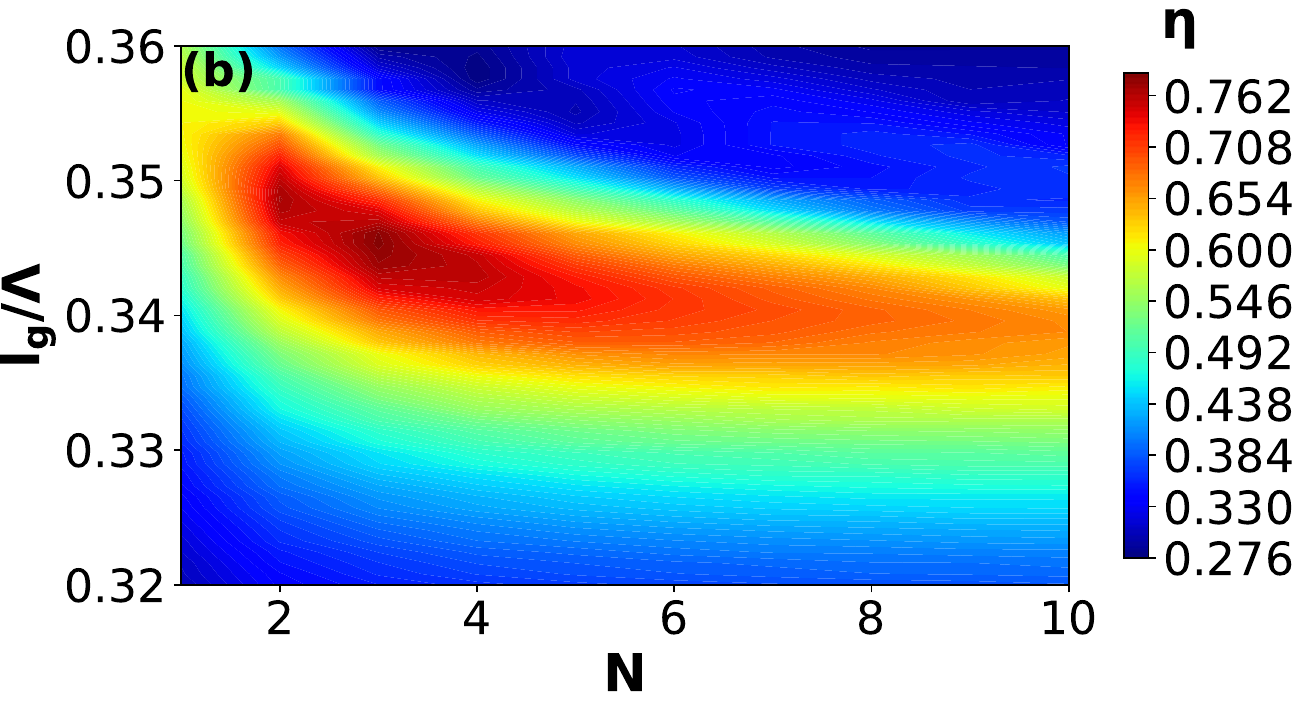}
    \caption{Coupling coefficient $\eta$. (a): $\eta$  as a function of the normalized  core thickness $t_c/\lambda$ and the number $N$ of periods of the grating.   $\Lambda$  is chosen according to the phase matching condition ranging from 219 nm for $t_c/\lambda=0.15$ to 224 nm for $t_c/\lambda=0.35$, $d$ = 20 nm, $l_g/\Lambda= 0.33575$. (b): $\eta$ as a function of $l_g/\Lambda$ and $N$  the number  of grating periods  for $t_c= 0.18 \lambda$, $d=20$ nm.} 
\label{fig:field-coupling-eta-tc}   
\end{center}
\vspace{-0.5cm}
\end{figure}

The last parameter to be studied is the number of graphene patches used in the grating in order to define the best coupler. The coupling coefficient $\eta$ is studied in the top graph of Fig.~\ref{fig:field-coupling-eta-tc}, it exhibits strong values reaching approximatively  56\%  for a  core thickness $t_c$ of 0.17 $\lambda$  for $N=11$ graphene patches. The length of the coupler region is then less than 2.5 $\mu$m for a wavelength set to 12 $\mu$m. The high values obtained for $t_c/\lambda = 0.17$ are  expected from the previous paragraph and a local maximum can also be seen for  $t_c/\lambda = 0.25$ due to a \gui{quarter-wave} effect, with the $l_g/\Lambda $ ratio obtained from the previous simulations. 

Then, one may \gui{wonder} if the obtained $\eta$ values around 60\% are \gui{optimal} for \gui{this type of graphene couplers. A wider range of the parameter triplet $(l_g/\Lambda, t_c, N)$ was spanned in order to answer this question. The results are shown} in the bottom graph of Fig.~\ref{fig:field-coupling-eta-tc}. 
 A curved band of high coupling coefficients is \gui{obtained with} peak values reaching 79\% obtained for $t_c/\lambda=0.17$ and  78\% obtained for $t_c/\lambda=0.18$ in both cases with only 3 periods of the graphene grating meaning a  coupler as short as 660  nm but for an increased duty cycle of 0.345  (see Table~S1 in Supplement 1). This optimized value of the duty cycle is higher than the one found for higher $N$ where a small decrease of the optimal duty cycle is observed confirming the limit theoretical value of 1/3 obtained for an infinitely long grating.


The long graphene sheet configurations are studied with the same method. A 60 $\mu$m-long sheet is chosen hereafter. In this case,  the coupling coefficient in the middle of such a long sheet is larger for a device without any grating. For example, the coupling efficiency toward the GP mode of the graphene sheet is  6\% for a grating with $\Lambda=220$ nm, $l_g/\Lambda=0.34$, and  $N=8 $ while it reaches 9\% without the grating (See Supplement 1). These values \gui{are always much smaller} than \gui{those} obtained for short configurations.
 The key parameter for these long sheet \gui{configurations} is, as previously shown, the waveguide core thickness, with  its optimal value in the range  [0.16,0.19] $\mu$m. This is due to the fact that the initial coupling occurs toward the high loss GP mode of the sheet, as a result the coupled field in the sheet decays rapidly and after ten micrometers the field vanishes.  
Consequently, the field recorded in the sheet at longer distance comes from the hydrid mode that propagates mainly in the waveguide core and couples to the GP mode.  For larger core sizes, higher order hybrid modes are also involved making the oscillations more complex as expected from coupled mode theory. 







Using both rigorous numerical methods describing graphene  as a genuine 2D conductivity material and physical analysis, the properties of graphene-based compact coupler toward graphene plasmon mode have been studied. Thanks to these results, optimized couplers have been designed which ensure a coupling efficiency of more than 75\% with a coupler region shorter than 700 nm for an operating wavelength of 12 $\mu$m. 
It must be pointed out that the developed method can take into account not only the fundamental mode of the input waveguide as incident field but also any linear combination of \gui{its intrinsic modes} including leaky ones since the formulation is written for a  general incident electromagnetic field.
More importantly, this formalism and its implementation are not limited to 2D configurations as studied \gui{here, but} they are valid for 3D ones since they are based on the full system of Maxwell's equations described in a vector finite element method. Consequently, they pave the way to the study of finite size effect in such 3D couplers involving 2D-material patches and can allow a direct comparison with future experimental results.

\vspace{-0.25cm}
\subsection*{Disclosure} The authors declare no conflicts of interest.
\vspace{-0.25cm}
\subsection*{Supplemental document} See Supplement 1 for supporting content.



\newpage

\end{document}